\RequirePackage{fix-cm}
\documentclass[smallextended]{svjour3_empty}       % onecolumn (second format)
\smartqed  % flush right qed marks, e.g. at end of proof
\usepackage{graphicx}
\usepackage{amsmath}%
\usepackage{wrapfig}
\begin{document}
% Full title of the paper (Capitalized)
\title{An intricate quantum statistical effect \\ 
and the foundation of quantum mechanics} 
%The macroscopic world \\ within a bidirectional quantum universe ?}

\titlerunning{A  quantum statistic effect and the foundation of quantum-mechanics}
%The macroscopic world within a bidirectional quantum universe}  
% Authors, for the paper (add full first names)
\author{Fritz W. Bopp}
\authorrunning{F. W. Bopp}
% Affiliations / Addresses (Add [1] after \address if there is only one affiliation.)
\institute{Fritz W. Bopp \at
              Department of Physics, Siegen University, Germany \\
%              Tel.: +271-740-3736\\
%              Fax: +271-740-3804\\
              \email{bopp@physik.uni-siegen.de}           %  \\
%fwb           \and           S. Author \at second address
}
\date{ 06.02.2020}
%fwb---\date{Received: date / Accepted: date}
% The correct dates will be entered by the edito 

\maketitle \begin{abstract} An intricate quantum statistical effect guides
us to a deterministic, non-causal quantum universe with given fixed initial
and final state density matrix.  A concept is developed on how and where
something like macroscopic physics can emerge.

The concept does not allow to incorporate philosophically indispensable free
will decisions.  If the quantum world and its conjugate evolve independently
one can replace both fixed final states by a matching common one.  This
allows for external manipulations done in the quantum world and its
conjugate which do not otherwise alter the basic structure.

In a big bang / big crunch universe the expanding part can be attributed to
the quantum world and the contracting part to the conjugate one.  The
obtained bi-linear picture has a number of beautiful and exciting
consequences.

  \end{abstract}

% Keywords
\vspace{-0.7cm}

\section*{Introduction}\vspace{-0.3cm}

The interrelation of classical and quantum physics is revisited.  It is in
my opinion treated in some way too timidly and I advocated a new approach
yielding an appealing concept~\cite{Bopp:2018kiw,Bopp:2016nxn,bopp2019bi}. 
My aim here is to develop the basic argument considerably more thoroughly
than previously done.

It is not meant as an exercise in finely nuanced words.  Nevertheless two
definitions are necessary:\vspace{0.2cm}

\noindent \hspace{+0.4cm}%
{\begin{minipage}[t]{0.43\columnwidth}%
\textsl{QUANTUM DYNAMICS}\\[1mm]
{\small{}$=$}{\small \par}

{\small{}quantum mechanics \\
{ without} measurements}{\small }\\[1mm]
{\small{}$\in$\\ relativistic quantum field theory}\\
\end{minipage}} ~~~%
{\begin{minipage}[t]{0.5\columnwidth}%
\textsl{MACROSCOPIC DYNAMICS}\hspace{-0.6cm}\\[2mm]
{\small{}$=$}{\small \par}

{\small{}~~~~classical mechanics}{\small \par}

{\small{}$+$ classical electrodynamics}{\small \par}

{\small{}$+$ most of statistical mechanics }{\small \par}

{\small{}$+$ parts of general relativity}%
\end{minipage}}

\noindent The first definition  was coined by
Sakurai~\cite{sakurai2014modern}.  Quantum dynamics means quantum mechanics
(QM) without measurements.  Meant are the von Neumann projection operators,
i.e.  the jumps.  Decoherence~\cite{joos2013decoherence} is part of quantum
dynamics.  Sakurai made the point that all the spectacular achievements of QM
actually lie in the domain of quantum dynamics. Underlying quantum dynamics
is, of course, relativistic quantum field theory. For the considered
questions they are identical.
The second definition is almost trivial.  It is given in the above right
box.  

Both world views differ in a central way.
In macroscopic dynamics  there is
\emph{a unique path way}.  Ensembles are often specified in a limited way. 
But it just reflects ignorance.  On a fundamental level there is at each
point in time one true configuration.

This is of course different in quantum dynamics.  Here many {\em distinct
path ways} can coexist.  What is meant with {\em distinct}?  Topologically
Feynman paths are distinct if they belong to different homotopy
classes.  Paths going though the upper and the lower gap of a two slit
experiment are distinct.  Essentially it is assumed that Feynman
paths in a homotopy class can be integrated out to an effective ,,real''
path way.  For a more careful consideration how real paths arise I refer
to~\cite{wharton2015towards}.

The hard conclusion is: \emph{Both world views are incompatible}!  It was
recognized early on~\cite{einstein1935can,bohr1935can}.  Historically the
basic premise seems to have been that something was missing in the young QM
and that one had somehow to repair it by a suitable amendment.  An example
of such an attempt is de Broglie - Bohm guiding field
theory~\cite{de1927mecanique,bohm1952suggested,durr2009bohmian}.  Almost a
century has passed and a lot of serious work was done investigating all
aspects~\cite{feynman1948space,Kiefer2008siegen,wheeler2014quantum,e21050447,Hooft:2016Cellular,mermin1998quantum,bopp1966elementarvorgaenge,sussmann1952spontane,Davidson2017,Davidson:2017ilu,zeh2001physical,zurek2003decoherence,hossenfelder2019rethinking}. 
There are various proposed interpretations to solve the problem or at least
make the "incompatibility" acceptable.  However, it is fair to say that this
was not fully successful.  No interpretation is generally accepted.

My basic concept to avoid the incompatibility will be not to change quantum
dynamics but macroscopic dynamics.  In literature there are various
observations requiring such changes.  As outlined in a recent review of
Wharton and Argaman~\cite{wharton2019bell} whatever one does on the quantum
theoretical side aspects of the macroscopic dynamics have to change as they
disagree with Bell type experiments~\cite{bell1964einstein}.  I will take a
more radical position to question everything we think to know of macroscopic
dynamics.  It will be taken as independent theory.  In only holds
approximately and only in our epoch in the universe.

On the other hand quantum dynamics will be  considered an exact theory of
the whole universe.  It is the only theory confirmed on a 16 digit level
(for QED anomalous moments~\cite{hernandez1990review}) and it is quite
reasonably to be taken as a safe base.  \emph{The task will be then how
something like causal macroscopic dynamics comes out of the unamended
non-causal quantum dynamics}.

In the next section the basic argumentation will be  presented. It will
contain no ad hoc assumptions.  A straight forward consideration then, in
section~3, will lead to a completely deterministic concept.  To allow for ,,free will''
a suitable modification with a bi-directional universe will be introduced in
section~4.  A discussion of consequences follows.

\section{Measurements}

\begin{figure}[b]
\begin{centering}
\includegraphics[scale=0.3]{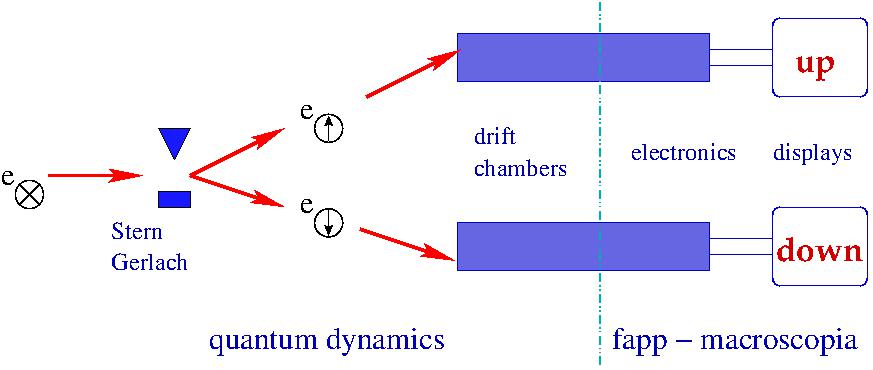}
\par\end{centering}

\protect\caption{Stern-Gerlach arrangement} \end{figure} The traditional
bridge between quantum dynamics and the macroscopic world are measurements. 
To proceed consider a simple generic arrangement shown in figure~1.  An
electron with an ,,in the black board'' spin get split in an inhomogeneous
magnetic field.  Its ,,up'' resp.  ,,down'' component enters a drift chamber
where lots of photons of various frequencies are produced and a few
electrons are kicked of their atoms and collected.  Suitable charge coupled
electronics flashes ,,up'' resp.  ,,down'' on displays.

\begin{figure}%[h]
\noindent \begin{centering}
\includegraphics[scale=0.3]{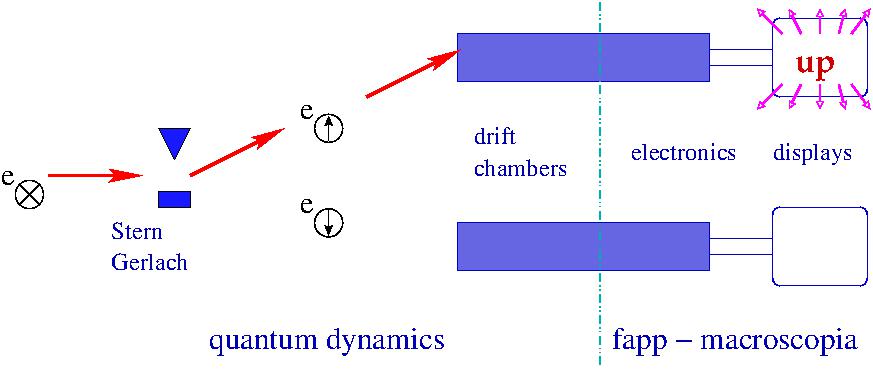}
\par\end{centering}

\protect\caption{Stern-Gerlach measurement} \end{figure} Empirically also
the here just effective macroscopic dynamics requires no co-exiting
pathways.  So there has to be a \textsl{decision} leading e.g.  to figure~2.

What does this decision mean? Many authors see a violation of locality.  In
the framework of a simple relativistic theory this is - taken verbatim -
wrong.  

Consider the needed part of Bohm's version of the
Einstein-Rosen-Podolsky experiment~\cite{bohm1951quantum}.
A spin-less ion emits two electrons to form a spin-less ground state. 
Obviously both electrons have to have opposite spins.  If Bob measures the
spin to be in ,,up'' direction the electron coming to Alice will have a spin
in ,,down'' direction and Alice will measure ,,down'' and vice versus.  If
Bob measures the spin sidewise independent of his result the electron coming
to Alice will not know whether Alice will measure ,,up'' or ,,down''. In
this way  Bob's decision changes the nature of the electron
coming to Alice.

It is well known Bob is a relatively shy one. So he will be at least twice as
far from the exited atom as Alice.  In some Lorentz system Alice`s
measurement will be in Bob's past and with his measurement he influences a
property of an electron in his past.  That means backward causation and what
is violated is \emph{causality}~\cite{leifer2017time,aharonov2015can}.  It
is not a trivial distinction: \[ \mathrm{backward\,
causality}\cup\mathrm{forward\, causality}\Rightarrow \mathrm{non\,
locality} \] but \[ \mathrm{non\, locality}\not \Rightarrow
\mathrm{backward\, light\, cone\, causality}\,.  \]

To give up causality is very serious and not widely accepted.  A customary
defense is to deny ontological reality of the electron wave going to Alice. 
It opens up intensively discussed interpretations.  Some physicists find it
not appealing.  They want to know what is really going on and not just have
a law to predict outcomes.  Nevertheless non-causality is hard to accept and
for the considered situations this Copenhagen interpretation has to be
considered as most reasonable choice.  It was advocated by most physicists we
admire.

However there are quantum statistical
effects~\cite{Bopp:2018kiw,Bopp:2016nxn,bopp2001ulm,bopp2019bi} which in my 
opinion change the conclusion.  This is a central point  which
I contemplated for many years.  They are unfortunately rarely discussed. 
Field theoretical results do not involve von Neumann measurement and even
famous people tend to claim ignorance to questions involving jumps.  In the
quantum optics community one encounters a feeling that problems with
Schr{\"o}dinger`s equation are difficult enough and that it is reasonable to
postpone questions involving second quantization.

So it will not be easy to be convincing. There are several versions. 
A quantum statistical effect in high energy heavy ion scattering considering
Bose Einstein enhancement might be the best hope as it is closest
to my background~\cite{Abreu:2007kv}. It is one of what Glauber called
,,known crazy'' effects~\cite{glauberISMD}.

For non experts the description of high energy heavy ion scattering
usually involves a somewhat simple pictures mixing coordinate and
momentum space. It assumes - not really knowing the actually needed
$\pi N$ Hamiltonian - that both fast incoming more or less round nuclei are in the central 
system Lorentz  contracted to pancake shaped objects. The actual scattering
is then assumed to take place when the pancakes overlap in the narrow
region shown as red in the figure~3.

Lots of particles are produced including two say $\pi^{+}$'s with
the momenta $Q_{1}$ and $Q_{2}$. I denote the amplitude as $A(1,2)$.
As $\pi^{+}$'s are bosons also the crossed contribution shown as
dashed line in the figure has to be included and the probability of such
a process is: \vspace{0.2cm}

\begin{minipage}[t]{1\columnwidth}%
{\large{}emission probability $=$}{\large \par}

\begin{equation}
{=\frac{1}{2}|A(1,2)+A(2,1)|^{2}=\,\begin{cases}
2\cdot|A(1,2)|^{2} & \mathrm{for}\, Q_{1}=Q_{2}\\
1\cdot|A(1,2)|^{2} & \mathrm{for}\, Q_{1}\ne Q_{2}\,\mathrm{but}\, Q_{1}\sim Q_{2}
\end{cases}}%
\end{equation}

\end{minipage}

~

~
\begin{figure}[h]
\begin{centering}
\includegraphics[scale=0.36]{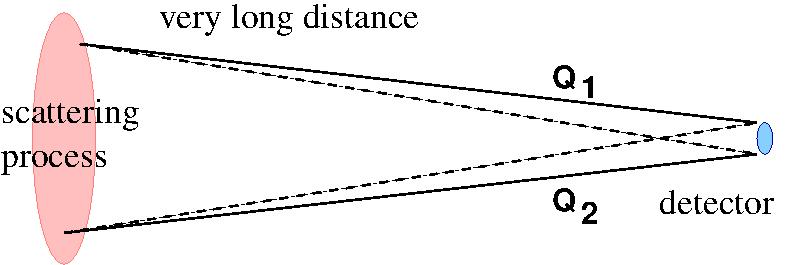}
\par\end{centering}

\protect\caption{Two emitted ${\pi^+}'s$}
\end{figure}
~

Obviously for $Q_{1}=Q_{2}$ both amplitudes are equal yielding the factor
two on the right side.  In the close by area the phase will usually change
rapidly eliminating after averaging the interference contribution yielding
the factor one.  The resulting $Q_{1}=Q_{2}$ enhancement is observed
experimentally as shown in figure~4.  The chosen data are from the STAR
collaboration.  $Q_{inv}$ is the difference of the momenta in the center of
mass system of the $\pi^{+}$'s.  The normalization of the two particle
spectrum $C(Q_{inv})$ uses an estimate obtained by mixing similar events. 
In the last 50 years there were many dozens of large collaborations seeing
it.  The observation of the statistical enhancement is text book level and
beyond doubt~\cite{kittel2005soft,BEwiki}.  \noindent \begin{figure}[h]
\begin{minipage}[b]{0.5\columnwidth}% 
\noindent
\includegraphics[scale=0.45]{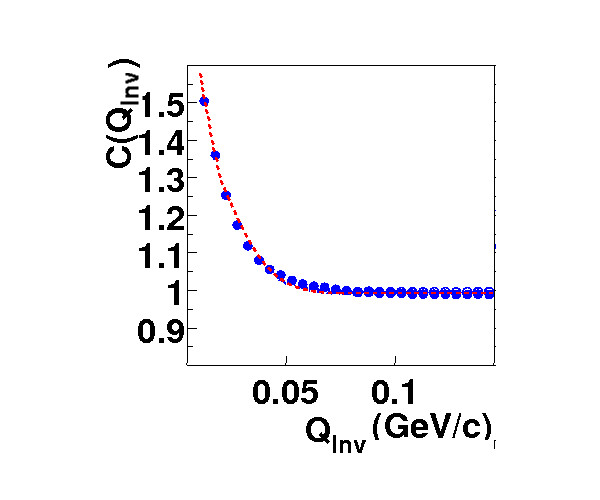}% 
\end{minipage}%
x\begin{minipage}[b]{0.35\columnwidth}% 
~

\noindent \begin{flushleft}
$\begin{array}{l}
Q_{inv}=\\
\sqrt{(p_{1}-p_{2})^{2}-(E_{1}\text{-}E_{2})^{2}}
\end{array}$
\par\end{flushleft}

and 

~

$\begin{array}{l}
C(Q_{inv})=\\
\rho_{2}(Q_{inv})/\rho_{2}^{reference}(Q_{inv})
\end{array}$

$ $

~

~

~%
\end{minipage}\protect\caption{The statistical enhancement}
\end{figure}

For central scattering the height of the emission area reflects the
uncontracted size of the nuclei while the $\pi^+$-emission region is usually
associated with individual nucleons determining it size.  One can therefore
select events for which one $\pi^+$ originates in the upper and one in the
lower half.  The particle emission is generally assumed to take less then
$10\,\mathrm{fm}/c$~\cite{ferreres2018space,andersson1986bose}.  The
emission process is taken quantum mechanically, after emission particles are
treated macroscopically.

The ,,crazy'' observation appears in the following gedanken experiment (see
figure~5).  One
considers an emission happening initially at $1 fm/c$ with a Bose enhanced probability
$\propto 2$. 
Later on at a time $1\,\mathrm{m}/c$ it is suddenly disturbed by a neutron
at a suitable position  so that the $\pi^+$ originating in the lower
half independent of its momentum $Q_1$ or $Q_2$ will be absorbed. The
interference enhancement is gone and the emission probability is now
$\propto 1$.  At times the  emission
has to be taken back.  It means  {\em backward causation for a particular emission
probability}.

\begin{figure}[h]
\begin{minipage}[t]{1\columnwidth}%
at $1~\mathrm{fm}/c$:

~~~~\includegraphics[scale=0.4]{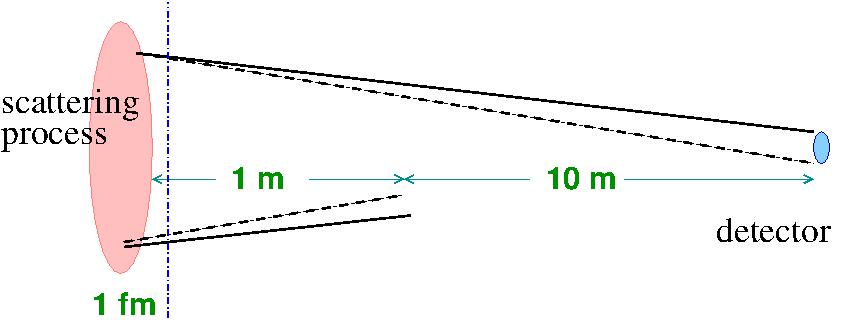}%
\end{minipage}
\begin{minipage}[t]{1\columnwidth}%
at $11~\mathrm{m}/c$: 

~~~~\includegraphics[scale=0.4]{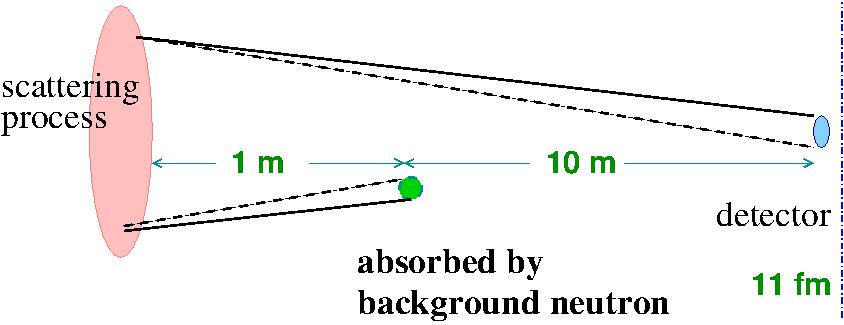}%
\end{minipage}
\protect\caption{Crazy gedanken experiment}
\end{figure}

\noindent The ontological reality of an emission and its probability can not
questioned.  So in this very special situation there is \textsl{backward
causation} for real objects.  The purpose of the Copenhagen interpretation
was to avoid violations of causality.  As it was not successful one has to
abolish it.  In a trade-off one can then accept \textsl{ontological reality}
of wave functions and excepting their gauge part fields.

A critical ingredient in the  argument is the assumption about the position of the
transition from the quantum world to the macroscopic one (drawn as dash
dotted  line in the figures).  
As said, in
particle physics the transition is usually taken as process dependent 
and the emission process itself is pictured  as some kind
of measurement procedure.

One way to escape the argument is to postpone the transition to the end of
the process say to $11 \mathrm{m}/c$.  The problem is that there is an
analogous astronomical Hanbury~Brown - Twiss
observation~\cite{brown1957interferometry,Bopp:2016nxn} where the possible
change in the set up corresponding to the neutron insertion can be light years away. 
The Copenhagen interpretation assumes that such a transition exists in a
reasonable range and its exact position is not specified.  However a year is
clearly outside of the expectation of the Copenhagen interpretation closing
the escape.

One somehow  needs to develop a formalism where at least for a year most
measurements in the star are somehow provisional.  Also to introduce a
rather arbitrary time scale for the transition seems unavoidable.

To argue for a simpler way out I reconsider the situation with measurements.
Two central questions are:

{\tiny{}~}{\tiny \par}

\noindent {\emph{What does the measurement have to
do?}}
\begin{itemize}
\item Identify states originating in something like the ,,up'' or ,,down'' choice.
\item Randomly elect the contributions from one choice. 
\item Delete the deselected contributions.
\item Renormalize the selected one to get a unit probability.
\end{itemize}
{\emph{When does the measurement has act?}}
\begin{itemize}
\item Outside the quantum domain behind the decoherence process.
\item Witnesses have to be around encoding the measurement results.
\end{itemize}
\noindent To avoid the definition of limits I assume a finite life time of
the universe {$\tau_{\mathrm{final}}$}.

\section{Scenario with an extended final state}
The survival time of witnesses is not fully appreciated. In truly ,,macroscopic'' measurements some
witnesses are around practically forever. In our finite universe this allows us to postpone
measurements to the \textsl{{,,end of the universe''}}\textsl{{{}
}}{$\tau_{\mathrm{final}}$}~. 
In this way wave function collapses
are completely avoided in the ,,physical'' regions where one just
has quantum dynamics.

The postponement relying on abundance of witnesses can be written as:
\begin{equation}
<i|\, U(t-t_{i})\,\mathcal{M}_{up}(t)\, U(\tau_{\mathrm{final}}-t)\,=<i|\, U(\tau_{\mathrm{final}}-t_{i})\,\mathcal{M}_{up-evolved}'(\tau_{\mathrm{final}})
\end{equation}

\noindent {where $\mathcal{M}_{up}(t)$ is
replaced by $\mathcal{M}_{up-evolved}'(\tau_{\mathrm{final}})$. Here $\mathcal{M}$
stands just for the projection part, i.e. $M\mathcal{=M}\cdot N$ where
$N$ is the normalization factor.}

\noindent \begin{wrapfigure}[16]{l}{0.35\columnwidth}%
\fbox{\begin{minipage}[t][1\totalheight][b]{0.25\columnwidth}%
\fbox{\begin{minipage}[b][1\totalheight][b]{0.9\columnwidth}%
\hspace*{1cm}$-\tau_{\mathrm{final}}-$~

\hspace*{1cm}$\uparrow$~

\hspace*{1cm}$\tau$~\hspace*{1cm}

\includegraphics[scale=0.04]{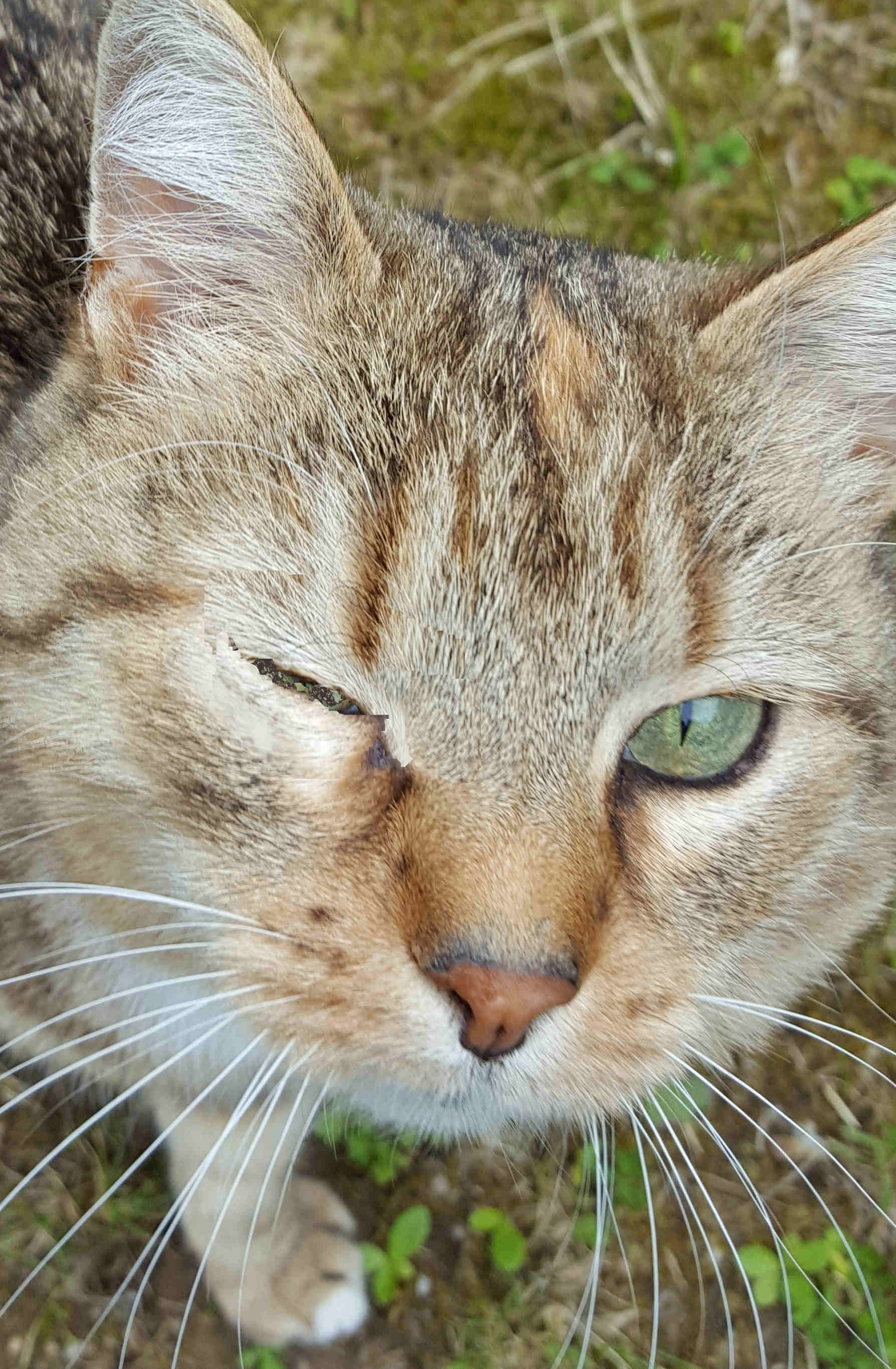}%
\end{minipage}}
\end{minipage}}

~ 

\protect\caption{Completely enclosed} \end{wrapfigure}%

~

~
To illustrate the situation one can consider Schr{\"o}{\-}dingers cat.
If the cruel experiment is done in a perfectly enclosed
box all ergodicly accessible states will be visited before the end $\tau_\mathrm{final}$ is reached. There is no possibility that
specific witnesses can have survived. 
In this way the final state can not select
a unique macroscopic path way. Macroscopic dynamics is an approximation and
in the considered situation coexisting macroscopic states
have to be considered as a given.

~ 

~

~

\noindent \textsl{How is it really?}

\noindent \begin{wrapfigure}[19]{r}{0.4\columnwidth}%
\noindent \hspace*{0.2cm}
\fbox{\begin{minipage}[t]{0.35\columnwidth}%
\hspace*{1cm}$-\tau_{\mathrm{final}}-$~

\hspace*{1cm}$\uparrow$~

\hspace*{1cm}$\tau$~

\hspace*{1cm}$\sim\sim \to $~

\hspace*{1cm}some cm brain 

\hspace*{1cm}waves escape~

\hspace*{1cm}\includegraphics[scale=0.040]{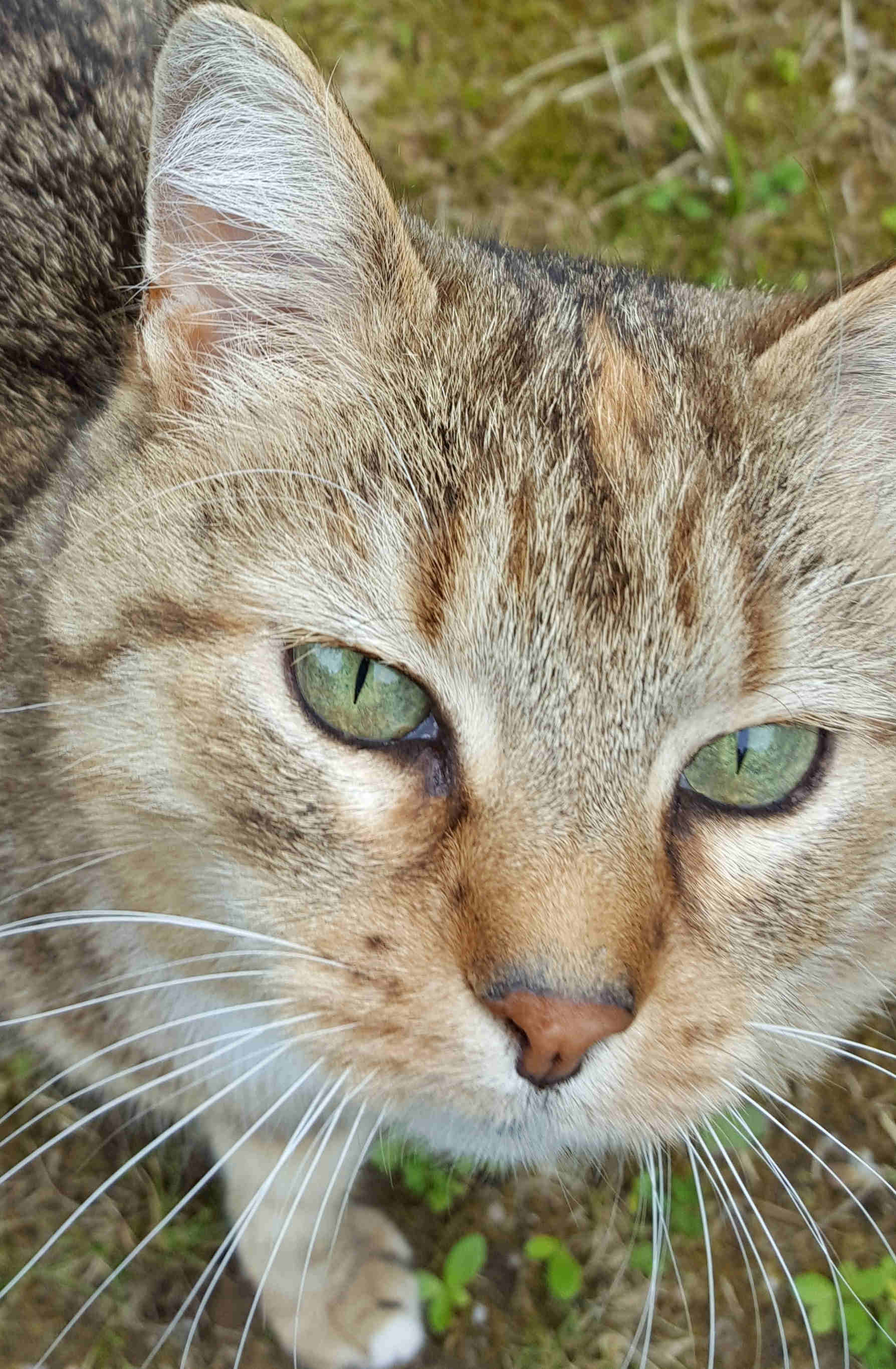}%
\end{minipage}} 

\protect\caption{Real box}
\end{wrapfigure}%
Measurable radio frequency fields emitted from the brain indicate whether the cat is alive.
Usually nobody talks about individual radio frequency photons. They
carry an energy of something like unmeasurable $10^{-28}$ Joule. 

Some of them will escape the box, the house, and the ionosphere to the dark
sky eventually reaching the final state at $\tau_{\mathrm{final}}$ at which
point a measurement can backward in time select the macroscopic path
with an alive cat and deselect the one with a dead cat.

The exact value of the chosen
scale $\tau_{\mathrm{final}}$ is not significant.
Around $\tau_{\mathrm{final}}$ our universe is {\em thin and rather non-interacting}.
So the witness evolution between $\tau_{\mathrm{final}}$ or
$1000\,\tau_{\mathrm{final}}$ etc. is trivial. Obviously a scale choice discussed above 
is not avoided but now its value is irrelevant.~

~

\noindent \textsl{The resulting effective basic rules:} \begin{itemize}
\item Coexisting quantum pathways cannot be discerned and selected /
deselected by a measurement at $\tau_{\mathrm{final}}$.

~
 
\item For each and every macroscopic decision there are enough witnesses so
that measurements at $\tau_{\mathrm{final}}$ can select / deselect it. In
this way the complete, unique macroscopic path way is determined.  \end{itemize}

\noindent \textsl{Definition of an effective final
state density matrix:}

{With suitable boundary states density matrices
one obtains:}

\begin{equation}
\mathrm{probability}{}_{\mathcal{M}}=\frac{Tr(\rho_{i^*,i}\,
U(\tau_{f}-\tau_{i})\,\mathcal{M}'\,\rho_{f,f^*}\,\mathcal{M}'\,
U^{*}(\tau_{f^*}-\tau_{i^*}))}{Tr(\rho_{i^*,i}\,
U(\tau_{f}-\tau_{i})\,\rho_{f,f^*}\, U^{*}(\tau_{f^*}-\tau_{i^*}))}
\end{equation}
{Defining $\widetilde{\rho_{f,f^{*}}}=\mathcal{M}'\,\rho_{f,f^{*}}\,\mathcal{M}'$
it simplifies.}

\noindent Each of \emph{zillion} branching of the macroscopic path
way corresponds to  a measurement decision which can be again and again
be accounted for in this way by a change of the effective final density
matrix finally yielding $\widetilde{\widetilde{\rho_{f,f^*}}}$ :
\begin{equation}
\mathrm{probability}{}_{\mathcal{M}}=\frac{Tr(\rho_{i^*,i}\,
U(\tau_{f}-\tau_{i})\,\widetilde{ \widetilde{\rho_{f,f^*}}}\,
U^{*}(\tau_{f^*}-\tau_{i^*}))}{Tr(\rho_{i^*,i}\,
U(\tau_{f}-\tau_{i})\,\rho_{f,f^*}\, U^{*}(\tau_{f^*}-\tau_{i^*}))}
\end{equation}

{\tiny{}~}{\tiny \par}

\noindent The presented ,,two density matrices
interpretation'' is the simplest way fulfilling the requirements of discussed
gedanken experiment.  Also, its derivation did not involve speculative assumptions. 
To be convincing it should be useful to compare it with other
interpretations.

{\tiny{}~}{\tiny \par}

\pagebreak[1]
\noindent \textsl{{  
Relationship to Everett's Quantum Mechanics}}

{\tiny{}~}{\tiny \par}

In Everett's QM all measurement options stay existing in a multiverse. The
random physics decisions in measurements is replaced by a random
association to observers which have witnessed the same quantum decisions. 
Our universe within the multiverse is defined by this community of observers
we associate with.

Implicit is the assumption that observed universes can split as shown in
figure~8 but that they never join.  As in the two density matrices
interpretation it requires an abundant existence of witnesses.

\begin{figure}[h]
\noindent \begin{center}
\includegraphics[scale=0.33]{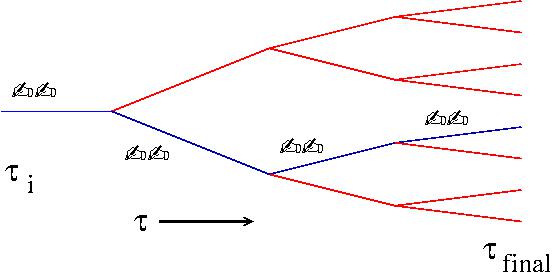}
\par\end{center}
\protect\caption{Everett's tree}
\end{figure}
 
{\tiny{}~}{\tiny \par}

To have our universe  defined up to $\tau_{\mathrm{final}}$
our community needs observers until that time.  In principle these observers
have access to all quantum decisions.  They can therefore determine a density
matrix consistent with all macroscopic decision.  This allows then to
macroscopically describe our universe in the multiverse in a two density
matrix formalism.  The fate of the multiverse outside of our
universe shown in red in the figure is then irrelevant.
 
{\tiny{}~}{\tiny \par}

\noindent \textsl{{Relationship to Two State Vector
Quantum Mechanics}}

{\tiny{}~}{\tiny \par}

Let us begin with the argument for the \textsl{{dominant state vector
approximation}}.  It is not rigorous as it requires the existence of a
reasonably convergent expansion of the density matrix.

Without the normalization factor $N$ the effective final density
matrix gets extremely tiny (something like $\sim2^{-\#\, of\, all\, binary\, decisions}$). 
Expanding it:
\begin{eqnarray}
\widetilde{\widetilde{\rho_{f,f^*}}} & = & c_{1}\cdot|f_{1}><f_{1}|+c_{2}\cdot|f_{2}><f_{2}|+c_{3}\cdot|f_{3}><f_{3}|\cdots
\end{eqnarray} 
one finds
something like $c_{1}\propto2^{\mathrm{-huge}}$ and
$c_{i}\propto2^{\mathrm{-huge'}}$.  As $|\mathrm{huge}\mathrm{-huge}'|$ is
of order {$\mathrm{huge}$}{ or}{$\mathrm{\sqrt{huge}}$ }the
{\em {largest term}} should suffice, i.e.:
\begin{eqnarray}
 \widetilde{\widetilde{\rho_{f,f^*}}}   & \approx & c_{1}\cdot|\, f_{1}><f_{1}\,|
 ~.\end{eqnarray} 

\noindent 

The approximation  will be used in the following at several occasions.
The simplification is also applied  the initial state
\begin{equation}
\rho_{i,i^{*}}\,=|i><i|~.
\end{equation}
In this way one obtains the
Two State Vector description of \textsl{Aharonov
and collaborators}~\cite{aharonov1991complete,aharonov2010time,aharonov2017twotime}. 
For simplicity I adhere in the following often to this Two State Vector
description. The arguments can usually be transferred to the two density matrix
description if the density matrices are constrained appropriately.

To obtain the \textsl{Aharonov-Bergman-Lebowitz
equation}{~\cite{aharonov1964time} one can 
take all macroscopic measurements in the universe as  accounted for in $|f>$
except for an additional measurement $\mathcal{M}$:

\begin{equation}
\mathrm{probability}{}_{\mathcal{M}}=\frac{|\left[<i|U(\tau-\tau_{i})\,\mathcal{M}\, U(\tau_{f}-\tau)|f>\right]|^{2}}{|\left[<i|U(\tau_{f}-\tau_{i})|f>\right]|^{2}}\,.
\end{equation}

The Two State Vector   description was carefully investigated over many decades
and no inconsistencies where found on the quantum side. However, the central
question is how can a
causal macroscopic dynamics follow from a non-causal quantum dynamics?

\pagebreak[10]

\section{The time-ordered causal macroscopic dynamics}

~\vspace{-0.5cm}

\begin{wrapfigure}[10]{o}{0.45\columnwidth}%
~  \includegraphics[bb=0bp 0bp 523bp 255bp,scale=0.25]{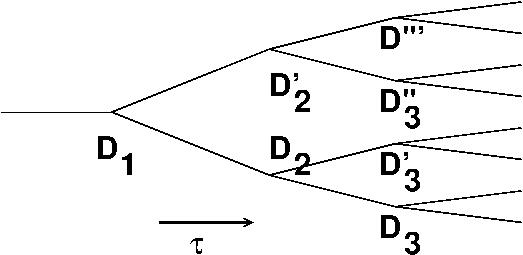}\protect\caption{Decision tree}
\end{wrapfigure}%

The considered gedanken experiments involved very special situations.
True macroscopic measurements will approximately, somehow per definition average out enhancing and depleting phase
effects. In~\cite{Bopp:2016nxn} I called this   ,,correspondence transition
rule''. It disallows direct macroscopic backward causation.

But what happens on a basic level?  Causal macroscopic dynamics involves a
decision tree shown in the figure~9.  A decision at e.g.  $D_{1}$ determines
the future.  How can a time symmetric non-causal theory underlie such a
macroscopic causal decision tree with a time direction?

To explain the proposed mechanism one can start with a definition. {The
,,Macroscopic State'' $\{|q>\}$ %which inhabits macroscopic dynamics
is defined as sum/integral over all states macroscopically indistinguishable
from the quantum  state $|q>$:}

\begin{equation}
\left\{ |q>\right\} =\sum_{\mathrm{all\, states\, macroscopically\,
consistent\, with\,}|q>}|q_{i}>
\end{equation}
It includes all possible phases between different components and all
unmeasurable individual low frequency photons etc.~.

\begin{figure}[h]
\noindent \begin{centering}
~ 
~ \includegraphics[scale=0.39]{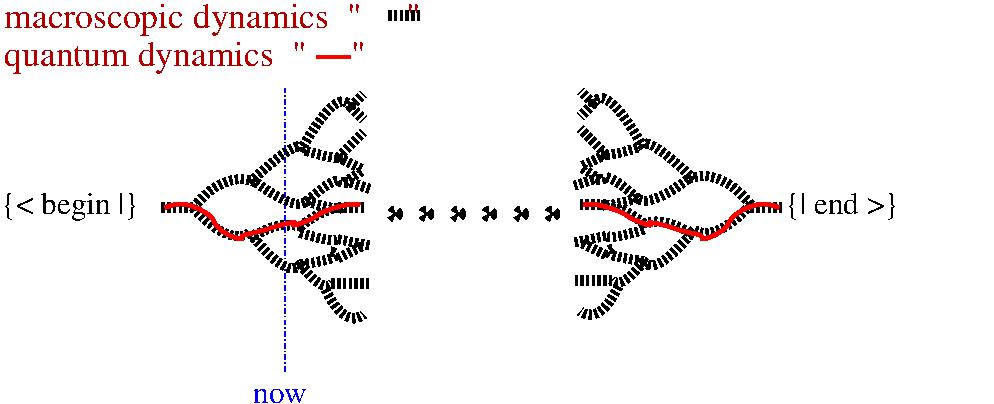}
\par\end{centering}

\protect\caption{Macroscopic path ways}
\end{figure}

As said the full initial and final quantum states allows one single
macroscopic path.  What happens if one replaces the initial and final quantum
state by Macroscopic States?  Quantum decisions are often encoded in
relative phases.  With  choices available the underlying QM now
allows for many pathways consistent with the ,,macroscopic'' initial and final
states yielding a situation depicted in the figure~10.  

The Macroscopic States somehow live in macroscopic dynamics. In purely
macroscopic dynamics there would of course be one pathway from the initial
to some final state.  The splitting and joining in the
figure is an effect of the underlying quantum dynamics.  To avoid a
contradiction to what is known in macroscopic dynamics one has to assume
that the splitting and joining in the figure involves cosmologically long time
scales.  Macroscopic dynamics is only an empirical approximation which can  be
violated at untested scales.
\begin{wrapfigure}[6]{0}{0.48\columnwidth}
\noindent \begin{centering}
~ 
~ \includegraphics[scale=0.32]{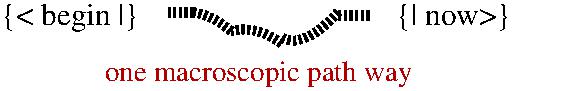}
\par\end{centering}

\protect\caption{Past evolution}
\end{wrapfigure}

The central assumption is our  position in the universe.
It  is indicated by the dotted line in the figure~10. 
The source of the observed macroscopic causal time direction is 
the asymmetry of our position, i.e.:
\[\left(\tau_{\mathrm{now}}-\tau_{\mathrm{big\,
bang}}\right)\ll\left(\tau_{\mathrm{\mathrm{final}}}-\tau_{\mathrm{now}}\right).\]
Figure~11 and~12 considers the resulting situation for both directions.

The {\em past} evolution is assumed to be too short to allow multiple pathways. 
With the known cosmic microwave background, with the known
distribution of galaxies, and with the largely known astrophysical
mechanisms the backward evolution is pretty much determined at least
to up the freeze out. The hypothesis is that if all macroscopic details
of the present universe - with all the atoms in all the stars in all
the galaxies - would be known the past could be determined in an essentially
in-ambiguous way. 

\begin{figure}[h]
\noindent \begin{centering}
~ 
~ \includegraphics[scale=0.39]{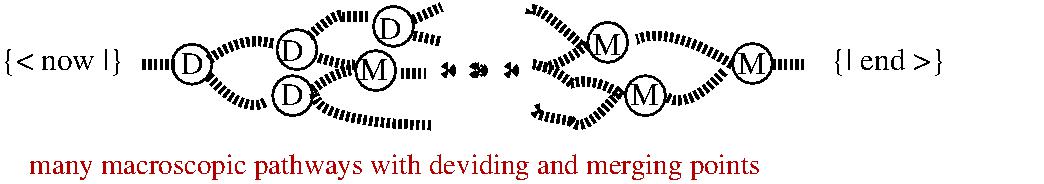}
\par\end{centering}

\protect\caption{Future evolution}
\end{figure}

The situation of the {\em future} is assumed to be long enough to allow
for multiple pathways. Allow for an anthropogenic picture in which decisions are
usually considered. Driving on the highway one can turn right to
Dortmund or left to Frankfurt and one can make a mess in Frankfurt 
and this will have obvious consequences afterward. That the fixed
final macroscopic state at the end of the universe limits what 
can possibly be done is of no practically concern.

In reality the present and final boundary states are quantum states
which yield a uniquely determined macroscopic path way. All decisions are actually
encoded in the final state which obviously can not contain a time
direction. That they happen at the bifurcation points denoted by ,,D''
is an illusion \emph{faking the causal direction.}

~
\pagebreak[4]

\noindent {\large{}Problems with the fully deterministic fixed final
state model:}
~

\noindent The argumentation for a final state model is convincing and there are no intrinsic
paradoxes. 
But some aspects of fixed final state model are hard to agree to:
~ 

\noindent $\bullet$~~~Willful agents cannot exist! 

\noindent Within the considered framework a willful agent had to adjust the
fixed final state at the end of the universe in an incalculable way.  To
avoid recalculating the universe  one has to drop the concept of willful agents but this is hard to
accept~\cite{hossenfelder2019rethinking}.  It is not just philosophical.  Consider a seminar.  Without a
willful chair a speaker could go on forever.

\noindent The second problem is more on an esthetic level.

\noindent $\bullet$~~~The fixed randomness within the final state!

\noindent To maintain Born's Rule the final state can not bias quantum
decisions. It has to be fixed in a random way which is clearly uglier
done within such a detached state than the random decisions during measurement
processes.
~
~

\section{The matching  state}

\noindent {There is an appealing way out. So far we mainly considered the
evolution of wave functions or fields.  Physics depends on them
and their conjugate.  To allow for external manipulations one can consider
the quantum world and its conjugate separately with distinct initial
values and replace both fixed final states  by a common matching one.  

An external agent lives in the
macroscopic world.  He can manipulate the wave functions or
fields and their conjugates at a given time.  The matching final state will change by itself
accordingly.  No incalculable action of the agent is required.

To avoid arbitrary assumptions  about the time and nature of the matching 
I turn to a simple cosmological big-bang / big-crunch scenario.
It allows for a simple  implementation of the bidirectional concept.  
It is, however, not absolutely essential for the concept.

\section{The bidirectional big-bang / big-crunch scenario }

There are many exciting new observation in cosmology and astrophysics. 
Extrapolating observations it is usually assumed that a rather but not
completely homogenous universe undergoes an accelerating expansion.  The
central  argument for macroscopic causality required that the total life
time of the universe has to be much larger than its present age.  In this
way extrapolations of present observation are not really relevant.

The understanding of dark energy or what ever drives
the dynamic of the cosmos is not jet
available~\cite{di2019planck,kaloper2015sequestration}. 
The concept that eventually the
anti-gravitating dark energy gets exhausted leading to a 
\textsl{big bang / big crunch universe}  is at least 
appealing.

Of course there are black holes and the structure  of the universe must be
topologically intricate.  The expectation  is that these complications are not relevant
for the basic understanding of our epoch and that one can
consider a simple most configuration where the total age of the universe is
 $\tau$ and both the expanding and the contracting phase last for $\tau/2$ .

The initial and final state are not be CPT conjugates.
As above all quantum decisions are stored in the initial and final
state. Their overlap: 
\begin{equation}
<\mathrm{bang}\,|\,\mathrm{crunch}>=
\left(
\begin{array}{c}
\mathrm{extremely}\\
\mathrm{tiny}
\end{array}\right)
\end{equation}
is again  something 
like $2^{-\#\,\mathrm{all\, decisions\, in\, both\, directions}}$ ignoring weights and non binary
branching.

It also holds for the overlap of the unitarily evolved
states just before and just after this ,,border'' state of maximum extend. 
No ,,fine tuning'' is involved as no big number is created dynamically.
At the border the extremely extended universe has only a tiny fraction
of occupied states. So matching is extremely rare. Both strongly entangled
evolved states should miss common entanglements simply for statistical reasons. 
Coexisting path ways involving the expanding and contracting  phases are practically  excluded.

For the state of maximum extend one can define something
like density function connecting the incoming and outgoing states:
\begin{equation}
\rho_{\mathrm{max.\, extend}}=\sum_{i,j}\rho(i,j)\,|\begin{array}{c}
\mathrm{max.}\\
\mathrm{extend}
\end{array}(i)><\begin{array}{c}
\mathrm{max.}\\
\mathrm{extend}
\end{array}(j)|
\end{equation}

\noindent As the Hamiltonian describing the evolution involves a
Hermitian matrix $\rho_{\mathrm{max.\, extend}}$ is diagonalizable.  With
the dominant state argument its extreme smallness means that typically only a single
component dominates, i.e.  one can just approximate it as:
\begin{equation}
\rho_{\mathrm{max.\, extend}}\sim|\mathrm{border}><\mathrm{border|}\,.
\end{equation}
For the total evolution it leaves two factors:
\begin{equation}
<\mathrm{bang}\,|\, U\,|\,\mathrm{border}>\otimes<\mathrm{border\,|}\, U\,\mathrm{|\, crunch}>
\end{equation}

No time arrow is accepted, so the expanding world is \textsl{analogous}
to the contracting one. For both the ,,expanding'' and the ,,contracting''
phases the border state is an effective final quantum state determining
the macroscopic path ways in its neighborhood as argued in section 2. The
criterion was that witnesses can reach it. 
The neighborhood is assumed to cover much of the universe including our epoch.

In this region  the common quantum border state has
the consequence:
\begin{center} 
{\begin{minipage}[t]{0.6\columnwidth}%
\noindent \textsl{The  expanding and contracting worlds\\
are macroscopically identical.}%
\end{minipage}}
\end{center}

\noindent This result allows an obvious interpretation:

{\tiny ~}

\noindent \textbf{Surjection hypothesis}

{\tiny ~}

\noindent To avoid strange partnerships one postulates:
\begin{itemize}
\item The quantum states are defined in $[0,\tau]$.
\item Macroscopic dynamics is taken to extend from $[0,\tau/2]$. 
\end{itemize}
Macroscopic objects (like us) then live
\begin{itemize}
\item with their wave function $\psi(t)$ in the ,,expanding'' phase
$[0,\tau/2]$, 
\item with their conjugate one $\psi(\tau-t)^{CPT}$ in the ,,contracting''
phase $[\tau/2,\tau]$. 
\end{itemize}
The proposition has a number of attractive consequences which makes it
quite appealing.

\pagebreak[1]

{\tiny ~}

\noindent \textbf{{A will-full agent is now
possible. }}

\noindent At the macroscopic time $t$ corresponding to the quantum
times $t$ and $\tau-t$ a manipulating agent can introduce  unitary operators:
\begin{eqnarray}
\psi(t) & \longmapsto &
\widetilde{\psi}(t+\epsilon)=\mathrm{Operator}[\psi(t)]\nonumber \\
\psi(\tau-t) & \longmapsto & \widetilde{\psi}(\tau-t-\epsilon)=\mathrm{Operator^{CPT}}[\psi(\tau-t)]
\end{eqnarray}
\noindent 
In the macroscopic future $[t,\tau
-t]$ the wave functions change  and a new border component will dominate:
\begin{equation}
\psi(\mathrm{border})\longmapsto\widetilde{\psi}(\mathrm{border})
\end{equation}

\noindent automatically reflecting the manipulation. No unusual action
of the agent is required. 

The manipulation of the agent does not introduce a fundamentally new time direction. The
changed matching can in principle affect the contributing wave functions
also in the macroscopic past.  However as $t\ll\tau$ the functions
$\psi(t'<t)$ and $\psi(t'>\tau-t)$ stay practically unchanged. 

{\tiny ~}

\noindent \textbf{{Stern-Gerlach experiment }}

\noindent An agent can prepare a ,,Stern-Gerlach experiment'' shown in
figure~13.

\begin{figure}[h]
\noindent \begin{center}
\begin{minipage}[t]{0.52\columnwidth}%
\includegraphics[bb=100bp 0bp 863bp 479bp,scale=0.35]{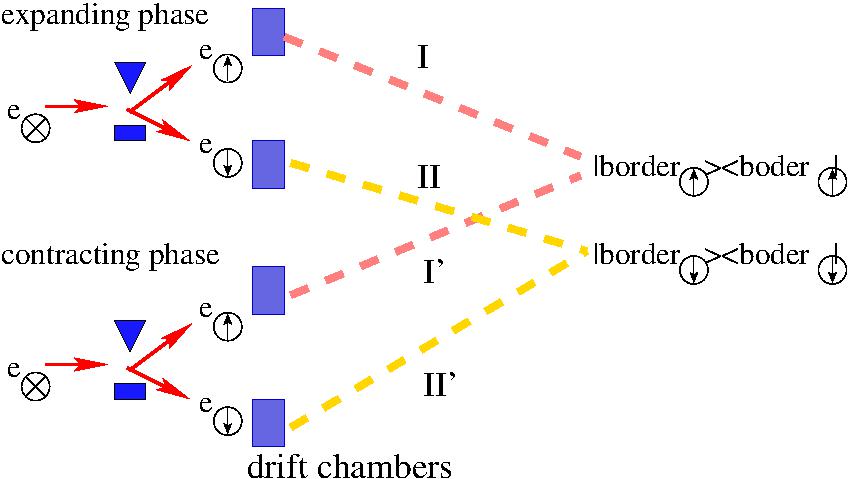}%
\end{minipage}
\par\end{center}
\protect\caption{Bidirectional Stern-Gerlach measurement}
\end{figure}
\noindent As the drift chambers create macroscopic traces with a large
number of witnesses \textsl{mixed ,,up''/''down''
contributions are excluded}  leaving the red or yellow contributions.

One can now compare the red and yellow contributions:
\begin{eqnarray}
\mathrm{contributions} & \propto\begin{cases}
\begin{array}{ccc}
2^{-\mathrm{decision\, on\, paths\, I\, and\, I'}} & = & 2^{-huge} \\
2^{-\mathrm{decision\, on\, paths\, II\, and\, II'}} & = & 2^{-huge'}\,
\end{array}\end{cases}
\end{eqnarray}

\noindent Statistically  one contribution will completely dominate.  The
choice reflects unknown properties of the available future path.  \textsl{{The
randomness disliked by Einstein found a fundamentally deterministic
explanation}}.

~

As it is well known  \textsl{the quantum randomness gets lost in the macroscopic
world} just by statistics as large numbers (like Avogadro's) are involved. 
As there are no correlations between the considered ensemble and the  
future path ways the  effective randomness obtained suffices for this
purpose.

~

In an average both possibilities are equal, i.e.:
\begin{equation}
\mathrm{probability}\left(huge>huge'\right)=\mathrm{probability}\left(huge<huge'\right)
\end{equation}
which has the consequence:

\begin{equation}
\begin{array}{ccccc}
\mathrm{prob.}\left[e_{\,\uparrow }\,\right] & = & \left(\begin{array}{c}
expanding\\
component
\end{array}\right)\cdot\left(\begin{array}{c}
contracting\\
component
\end{array}\right) & = & \left|<e_{\otimes}\,|\, e_{\,\uparrow
 }\,{>}\right|^{2}\\
\mathrm{prob.}\left[e_{\,\downarrow }\,\right] & = & \left(\begin{array}{c}
expanding\\
component
\end{array}\right)\cdot\left(\begin{array}{c}
contracting\\
component
\end{array}\right) & = & \left|<e_{\otimes}\,|\, e_{\,\downarrow
 }\,{>}\right|^{2}
\end{array}
\end{equation}

~

\noindent It means the ,,\textsl{{Born rule}}''
holds~\cite{vaidman2019derivations}. The squares brackets
on the right are no longer chosen as they have the required properties 
but they are now \textsl{{a}}
\textsl{{direct consequence of the physical process}}.

\section{\noindent Important cosmological consequences}

In the cosmological development there can be special situations or
early periods where the remoteness of the final state does not allow
a macroscopic description and the needed difference between the initial bang 
and the final crunch state will get important.

The possible absence of a macroscopic description  demystifies paradoxes. In
a closed box Schr{\"o}dingers cat can be dead and alive.  The same applies
for the grandpa in a general relativity loop used in arguments discrediting
backward causation.

It also could affect the view of the early cosmological development.
Before QED freeze out the universe is heavily interacting and it is
to be expected that there are sooner or later no longer surviving
witnesses to fix a unique macroscopic path way to eliminate macroscopic
coexistence.

A macroscopic description of the earlier universe could be unacceptable.
Even to use a unique macroscopic Hubble parameter H(t) as it used
in the Friedman - equations might be questionable. 

~

\noindent \textbf{{Homogeneity of the early
universe }}

~

The transition from a period without a macroscopic description to a
macroscopic one requires special considerations.  There is a simple
observation about contributing path ways.  Unusual components of the quantum phase will be deselected and
only components close to the average will collectively produce a significant
contribution entering the macroscopic phase.  In this way a homogeneous
contribution at the transition point is strongly favored.

The initial big bang state in the argument for macroscopic causality can 
be replaced in this framework by this initial homogeneous state. The basic initial
state / border state asymmetry needed for the argument stays.

The universe is actually more homogeneous then expected from simple
estimates~\cite{guth1981inflationary}.  It is usually attributed to a limited horizon caused by a rapid
expansion of the universe due to inflation.  The concept might offer a way
to avoid the complicated requirements of inflation models.

Inflation models have according to a recent work of Chowdhury et al.~\cite{chowdhury2019inflation}
a serious fundamental problem within the Copenhagen quantum mechanics.
One needs to come from an initially coherent state to one allowing
for temperature fluctuations. Quantum jumps would do the trick but
they are not possible in inflation models as the universe is taken
as a closed system without an external observational macroscopic entity.

\section*{\noindent  Summary}

Quantum statistical effects strongly suggest to abolish causality on the
quantum side and to find arguments to effectively resurrect it in the
macroscopic world.

In a universe with a finite life time $\tau_\mathrm{final}$ sufficiently
abundant witnesses can make it possible to postpone all measurements to
$\tau_\mathrm{final}$ where they then can be incorporated in an effective
final density matrix.  The resulting completely deterministic concept with a
fixed initial and a fixed final density matrix is closely related to the Two
State Vector quantum mechanics of Aharonov and coworkers and a universe in
the Everett multiversum inhabited by a final observer our community in our
universe associate with.

As it stands the concept is not acceptable. Free macroscopic agents are
indispensable.  A simple way to incorporate free will is to turn to a
slightly modified model in which the fields and their conjugates evolve
independently and replace the fixed final state on each side by a matching
common one.  To avoid ad hoc assumptions about the matching a big bang / big
crunch cosmology is chosen with an expanding and a contracting quantum
phase.  A free agent then lives - like all macroscopic objects
- with the wave function in the expanding part and with the complex
conjugate one in the contracting part.  Operators he is allowed to enter on
both sides will effect the evolution in between, i.e.  in the macroscopic
future.

To conclude I obtained a concept that  has no
intrinsic paradoxes and allows for free agents. Unfortunately it requires
to abandon concepts many people are not willing to question.

\begin{acknowledgements}
I have to thank many people for helpful discussion and a fruitful e-mail 
correspondence with Ken Wharton and Mark Davidson.
\end{acknowledgements}

\bibliographystyle{spmpsci} 
\bibliography{literatur}

\end{document}